\documentclass[apjl]{emulateapj}

\usepackage{natbib}

\shorttitle{A solar mass star with two cold giant planets}
\shortauthors{Beaulieu et al.}

\begin{document}


\title{Revisiting the microlensing event OGLE 2012-BLG-0026: A solar mass star with two cold giant planets}


\author{J.-P.~Beaulieu\altaffilmark{1,2}, D.P.~Bennett\altaffilmark{3,4}, V.~Batista\altaffilmark{1},   A.~Fukui\altaffilmark{5},  
J.-B.~Marquette\altaffilmark{1}, S.~Brillant\altaffilmark{6},  A.A.~Cole\altaffilmark{7}, L.A.~Rogers\altaffilmark{8,9}, T.~Sumi\altaffilmark{10}, 
F.~Abe\altaffilmark{11}, A.~Bhattacharya\altaffilmark{3,4}, N.~Koshimoto\altaffilmark{10}, D.~Suzuki\altaffilmark{3,4}, P.J.~Tristram\altaffilmark{12},  C.~Han\altaffilmark{13}, A.~Gould\altaffilmark{14}, R.~Pogge\altaffilmark{14},  J.~Yee\altaffilmark{15} }
\altaffiltext{1}{Sorbonne UniversitŽs, UPMC Univ Paris 6 et CNRS, UMR 7095, Institut dÕAstrophysique de Paris, 98 bis bd Arago, 75014 Paris, France; beaulieu@iap.fr; batista@iap.fr; marquett@iap.fr}
\altaffiltext{2}{LESIA Observatoire de Paris, Section de Meudon 5, place Jules Janssen 92195 Meudon, France}
\altaffiltext{3}{University of Notre Dame, Department of Physics, 225 Nieuwland Science Hall, Notre Dame, IN 46556-5670, USA; bennett@nd.edu,suzuki@nd.edu}
\altaffiltext{4}{Laboratory for Exoplanets and Stellar Astrophysics, NASA/Goddard Space Flight Center, Greenbelt, MD 20815, USA}
\altaffiltext{5}{Okayama Astrophysical Observatory, National Astronomical Observatory of Japan, Asakuchi, Okayama 719-0232, Japan; afukui@oao.nao.ac.jp}
\altaffiltext{6}{European Southern Observatory (ESO), Karl-Schwarzschildst. 2, D-85748 Garching, Germany;\\ sbrillan@eso.org}
\altaffiltext{7}{School of Physical Sciences, University of Tasmania, Private Bag 37 Hobart, Tasmania 7001 Australia; Andrew.Cole@utas.edu.au}
\altaffiltext{8}{Department of Astronomy and Astrophysics, 5640 S. Ellis Ave, Chicago, IL 60637, USA;  larogers@uchicago.edu}
\altaffiltext{9}{Sagan Fellow, Department of Earth and Planetary Science, University of California at Berkeley, 501 Campbell Hall 3411, Berkeley CA, 94720, USA}
\altaffiltext{10}{Department of Earth and Space Science, Graduate School of Science, Osaka University, Toyonaka, Osaka 560-0043, Japan; sumi@ess.sci.osaka-u.ac.jp}
\altaffiltext{11}{Institute for Space-Earth Environmental Research, Nagoya University, Nagoya University, Nagoya 464-8601, Japan}
\altaffiltext{12}{Mt. John University Observatory, P.O. Box 56, Lake Tekapo 8770, New Zealand}
\altaffiltext{13}{Department of Physics, Chungbuk National University, Cheogju 361-763, Republic of Korea; cheongho@astroph.chungbuk.ac.kr}
\altaffiltext{14}{Department of Astronomy, Ohio State University, 100 W 18th Ave, Columbus, OH 43210, USA;  gould@osu.edu;pogge.1@osu.edu}
\altaffiltext{15}{Sagan Fellow, Harvard-Smithsonian Center for Astrophysics, 60 Garden St, MS-15, Cambridge, MA 02138, USA;  jyee@cfa.harvard.edu}

\begin{abstract}
Two cold, gas giant planets orbiting a G-type main sequence star in the galactic disk have previously been discovered in the high magnification microlensing event 
OGLE-2012-BLG-0026 \citep{2013ApJ...762L..28H}. Here we present revised host star flux measurements and a refined model for the two-planet system using additional light curve data. 
We performed high angular resolution adaptive optics imaging with the Keck and Subaru telescopes
at two epochs while the source star was still amplified. We detected the lens flux, $H=16.39 \pm 0.08$.
The lens, a disk star, is brighter than predicted from the modeling in the original study. We revisited the light curve modeling
using additional photometric data from the B\&C telescope in New Zealand and CTIO 1.3m H band light curve. We then 
include the Keck and Subaru adaptive optic observation constraints.
The system is composed of a $\sim 4-9$ Gyr lens star of $\rm M_{lens} = 1.06 \pm 0.05~\,M_\odot$  at a distance of $\rm D_{lens} = 4.0 \pm 0.3~$kpc, 
orbited by two giant planets of $\rm 0.145 \pm 0.008\ M_{\rm Jup}$ and $0.86 \pm 0.06~\rm M_{\rm Jup}$ with projected separations of $4.0 \pm 0.5 $ AU and $4.8 \pm  0.7$ AU respectively. 
 Since the lens is brighter than the source star by $16  \pm 8 \%$ in H, with no other blend within one arcsec, it will be possible to estimate its 
 metallicity by subsequent IR spectroscopy with 8--10~m class telescopes. By adding a constraint on the metallicity it will be possible to refine the age of the system.
\end{abstract}


\section{Introduction}

Over the last 20 years, several methods probing different sections of the exoplanet zoo 
have been used: radial velocity, stellar transits, direct imaging, pulsar timing, transit timing,
astrometry and gravitational microlensing. These discoveries have already challenged
and revolutionised our theories of planet formation and dynamical evolution. Of the nearly 
2000 confirmed planets known to date, 488 of them are in multi-planet systems, including  
two unusual systems with two cold gas giant planets, discovered by microlensing. The first one (OGLE-2006-BLG-109Lb,c) is a half-scale model of our solar system \citep{2008Sci...319..927G, 2010ApJ...713..837B}. The second one, detected 
in the microlensing event OGLE-2012-BLG-0026, is composed 
of two giant planets orbiting a G-type main sequence star (\citet{2013ApJ...762L..28H}, H2013 hereafter).

H2013 identified four possible sets of system parameters as a consequence of the well known close/wide degeneracy in the lens equation. They estimated the physical parameters using a parallax constraint and a measurement of the Einstein ring radius. 
They also noted a significant contribution of blended (unmagnified) flux to the light curve and recognized that sub-arcsecond imaging is required in order to separate the source$+$lens from possible contamination by unrelated stars. This de-blending is critical in order to properly estimate the brightness of the lens; such a measurement can then be used to refine the light curve model and decrease uncertainties on the physical parameters of the system. Here
we follow the approach described in detail by \citet{2014ApJ...780...54B, 2015ApJ...808..170B}. 
We measure the lens flux in the microlensing event OGLE-2012-BLG-0026 using Keck and Subaru and compare it to the predictions of H2013 and stellar models. We revisit the modeling of OGLE-2012-BLG-0026 including more photometric data, update estimates of the source radius, and use the new flux constraints from Keck and Subaru to draw new conclusions about the physical parameters of the system.

\section{ Detecting light from the lens star }

In most cases it is possible to detect and to study (or to put upper limits on) the host lens stars with 
high angular resolution observations with adaptive optics such as Keck \citep{Sumietal2010,  2014ApJ...780...54B,	2015ApJ...808..170B},  VLT  \citep{2008ASPC..398..499D, 2010ApJ...711..731J, 2011A&A...529A.102B, 2012A&A...540A..78K}, Subaru \citep{2015ApJ...809...74F}, GEMINI, MAGELLAN, and space based observations with HST \citep{2007ApJ...660..781B, 2009ApJ...695..970D, 2015ApJ...808..169B}.
High angular resolution allows us to resolve the source star from its unrelated neighbors, while the source and lens stars will generally still be blended together. Indeed, at the time of the microlensing event, the lens star must be less than $\sim 1 $ mas away from the source. The relative proper motion being typically $\sim 5 $ mas yr$^{-1}$, it will require several years to detect a centroid shift 
of the blended lens and source \citep{2007ApJ...660..781B} and more than a decade to finally see  
the lens and the source well separated \citep{2015ApJ...808..170B, 2015ApJ...808..169B}. 

Fortunately, it is possible to derive strong constraints shortly after the end of the microlensing event. 
Indeed, the microlensing models  determine the H-band brightness of the source star, so it is usually possible to determine
the H-band brightness of the host star (lens) by subtracting the source flux from the high angular
resolution measurement of the combined host$+$source flux. 
This measurement can be used with a mass distance relation as in Equation 1 from \citet{2007ApJ...660..781B},  
and an H-band mass-luminosity relation to yield a unique solution for the host star mass. 
This would yield the planetary mass and star-planet
separation in physical units because the planet-star mass ratio and the separation in
Einstein radius units are already known from the microlensing light curve.

\subsection{Keck and Subaru Observations}

The field containing the source star OGLE-2012-BLG-0026 has been observed by the VISTA 4m telescope 
in JHK as part of the VVV survey \citep{2010NewA...15..433M} monitoring the disk and the bulge of our galaxy. 
 We developed a suite of tools using \texttt{astropy}\footnote{\texttt{http://www.astropy.org}} \citep{2013A&A...558A..33A}, \texttt{astroML} \citep{astroML}, \texttt{TOPCAT} \citep{2005ASPC..347...29T}, and the  \texttt{AstrOmatic} programs \texttt{SExtractor} \citep{1996A&AS..117..393B} and \texttt{PSFEx} \citep{2011ASPC..442..435B}.
We extracted JHK images centered on the target from the ESO archive, and performed PSF photometry.
We performed astrometric  and photometric calibrations using the 2MASS survey \citep{2006AJ....131.1163S}. 
The resulting catalogs will be used to calibrate the AO data. 


First, we used the Keck-II telescope on Mauna Kea, with the NIRC2 imager and laser guide star, at medium resolution (pixel scale of 0.02 arcsec and a field of view of 20 arcsec). We obtained 5 H-band images with an exposure time of 30 seconds each 
on May 6, 2012 (HJD=2456062.081), while the source star was still magnified by $A=1.76$. The individual exposures were obtained in a dithered pattern with an amplitude of 1 arcsec.
We observed a second epoch on July 28, 2012 (HJD=2456134.834) with the IRCS camera on Subaru while the source was amplified by  $A=1.20$.  We obtained 20 dithered exposures of 30 seconds each.

The Keck and Subaru observations were reduced with the same procedure described by \citet{2014ApJ...780...54B}. For each data set, we correct first for the dark current and the flat-fielding using standard procedures.  Using the catalogs we generated from VVV 
images we compute a first astrometric solution for each image. We then adopt one AO image as a reference, and build up a catalog of sources using \texttt{SExtractor}. We refined the astrometry of the other frames using this catalog. We visually inspect the individual images, then use \texttt{SWARP} \citep{2010ascl.soft10068B} to stack them. Figure~1 shows H-band images from VVV, Keck, and Subaru. We identify no bright companion within the PSF of the lensing survey telescopes, suggesting that the blended light observed by OGLE is actually lined up to better than 0.1 arcsec with the source. 

We measured the flux of all sources in the field using \texttt{SExtractor} with aperture photometry. We cross-matched the VVV, Keck and Subaru catalogs and determined the zero points. We then cross identified Keck and Subaru sources to double check the consistency of the zero points. We obtained the following measurements of OGLE-2012-BLG-0026:
$H_{\rm Keck}=15.43 \pm 0.05$ while the source was amplified by 1.76 and $H_{\rm Subaru}=15.60 \pm 0.05$ while the source was amplified by 1.20.
The measured FWHM are 130 mas for Keck and 170 mas for Subaru. There are no resolved blends contributing to the measured flux by the non-AO telescopes within  $\sim$ 1 arcsec. The chance alignment of a blend with the source and lens to within the angular scale of the AO measurements is far lower than 1\%, so we identify the origin of the blended light as the lens itself. 

\begin{figure}
\includegraphics[scale=0.33,angle=-90]{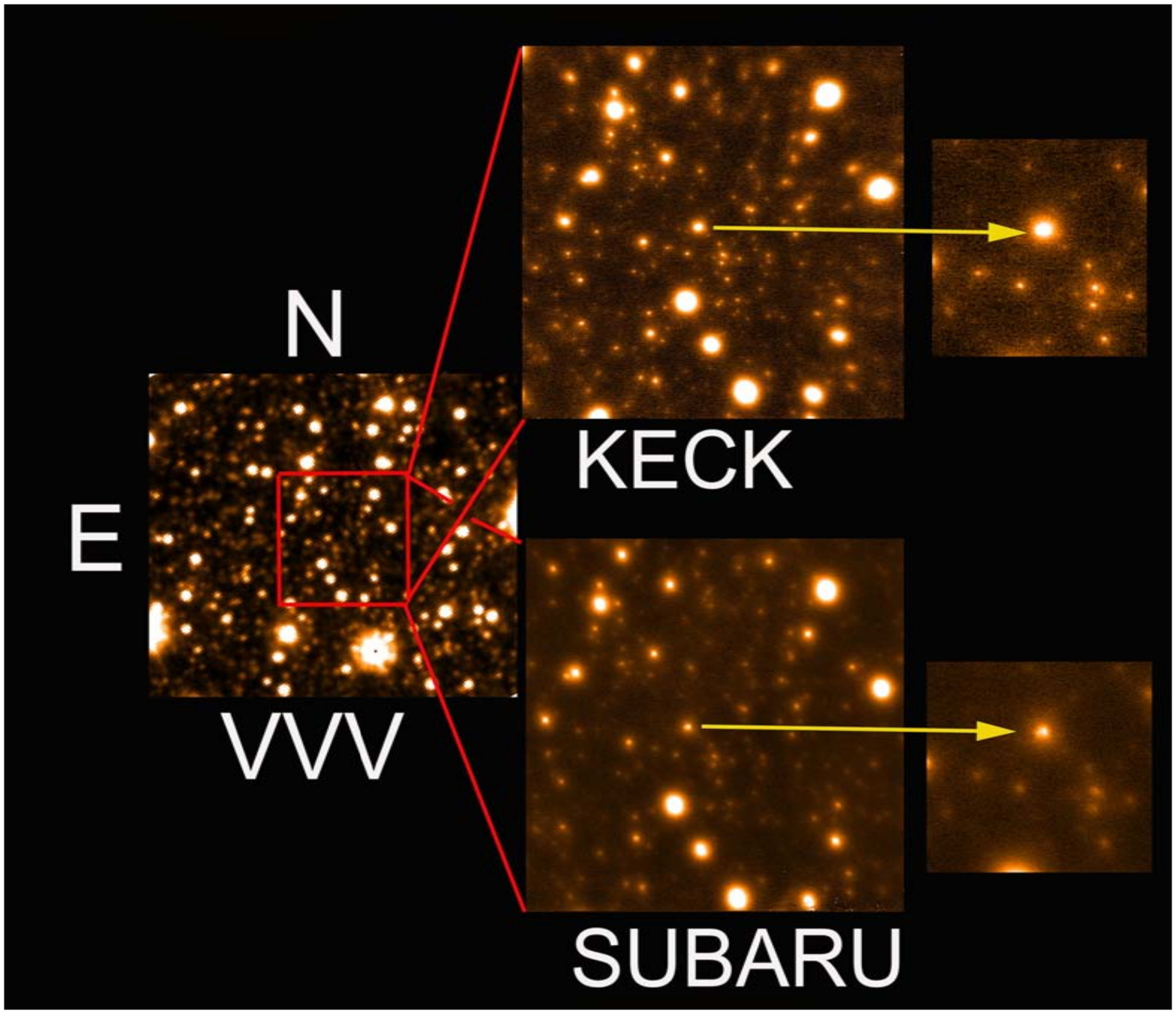}
\caption{H band images centred on the source of the microlensing event OGLE 2012-BLG-0026. 
The left insert is an H-band image obtained by the ESO VISTA 4m telescope as part of the VVV survey. The upper right panel has
been obtained with Keck, while the Subaru image is shown on the lower panel. They are both 18 arcsec square images. 
At the sub-arcsecond level, there are no bright  stars close to the source contributing significantly to the observed blended light in OGLE photometry (4 arcsec images). 
Any bright blend would have to be aligned with the source to better than 0.1 arcsec.\label{fig1}}
\end{figure}

\begin{figure}
\includegraphics[scale=0.45,angle=0]{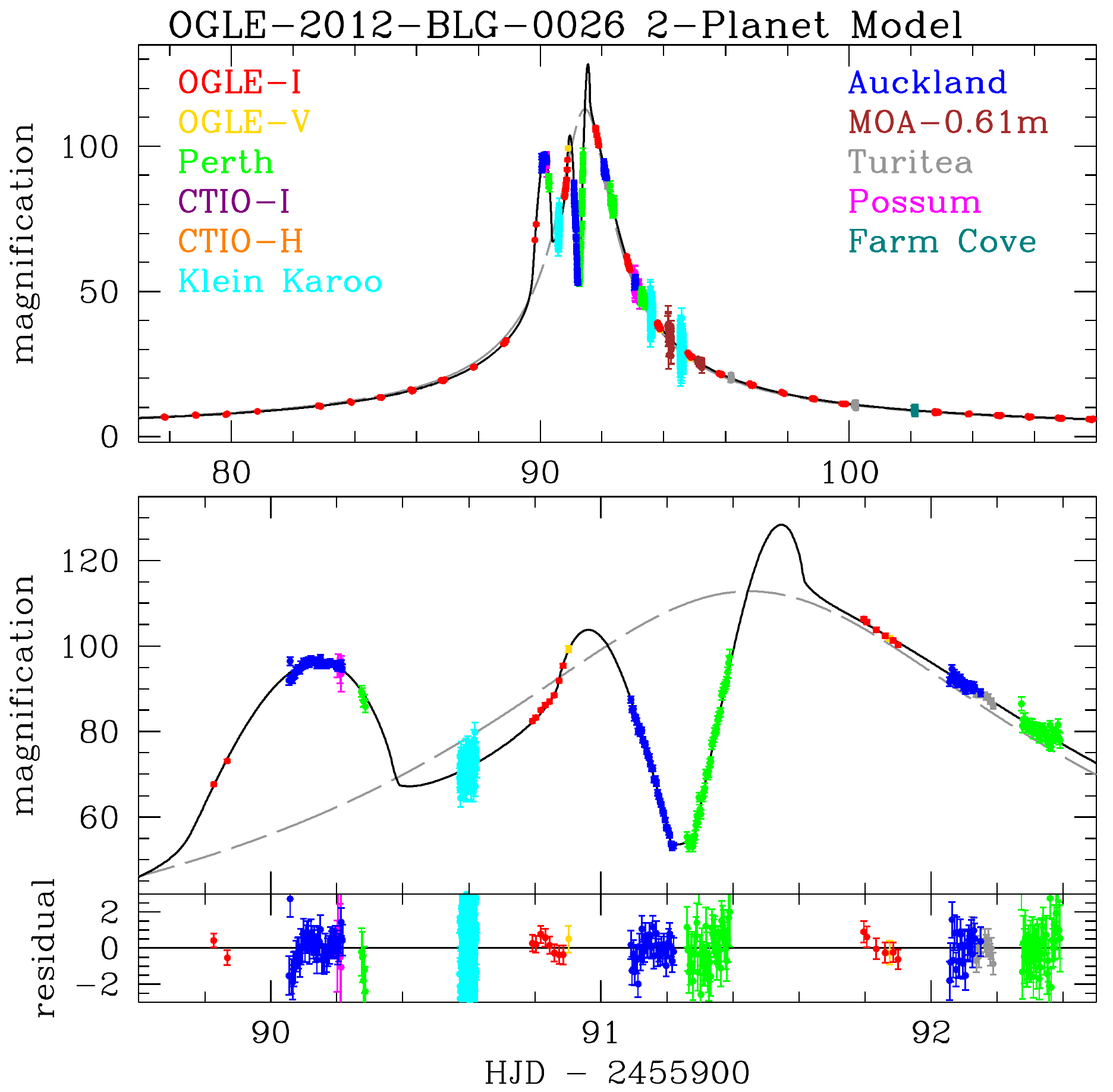}
\caption{The top panel shows the complete photometric light curve and best-fit, two-planet model of OGLE-2012-BLG-0026.  The 
different observatories are labeled and color-coded. The lower panel gives a zoom on the peak, the best model (solid), 
and the point source point lens model (dash). Below are plotted the residuals of the best fit model. 
We used the same datasets as H2013, with the addition of CTIO 1.3m H-band data, and I-band data from the MOA-0.61m. 
\label{fig2}}
\end{figure}

\subsection{Estimating extinction}
H2013 estimated the extinction towards the source to be $A_I=2.25$ by analyzing (V-I,~I) color magnitude diagrams and using the constraint on the red clump position.
This is compatible with the extinction maps provided by Gonzales et al., (2012) that  gives $A_I=2.27$ and $A_H=0.525 \pm 0.09$ with the \citet{2009ApJ...696.1407N} extinction law. 
Following \citet{2015ApJ...808..169B}, we adopt as a scale height of the dust towards the galactic bulge  $\rm \tau_{dust} = (0.120 kpc)/sin(b)$ (where b is the galactic latitude). 
The extinction to the lens $\rm A_{H_L}$ as a function of the extinction to the source $\rm A_{H_S}$ and distances to the source $D_S$ and to the lens $D_L$  reads: 
$$\rm A_{H_L}=(1-e^{ -{  D_L/ \tau_{dust}}  }) / (1-e^{ -{  D_S/ \tau_{dust}}  })  A_{H_S}$$
In the following, extinction to the lens will be estimated this way.

\subsection{Constraining the lens in H band}
An H-band light curve of the microlensing event was obtained by the ANDICAM camera mounted on the CTIO 1.3m telescope. Using the model described in H2013 and performing the $(I-H)$ regression, we can compute the baseline flux of the source in $H$ band to be  $H_{\rm S,~fitted}=16.56 \pm 0.01$. Using the Keck and Subaru measurements, we estimate the light coming from the lens to be $H_{\rm L,~Keck} = 16.48  \pm 0.13$ and $H_{\rm L,~Subaru} = 16.34 \pm 0.10$. 
The weighted average of these two measurements gives $H_{L} = 16.39  \pm 0.08$. 
The lens is brighter than the source by $16  \pm 8  \%$, slightly bluer than the source in $(V-I)$ by $\Delta (V-I) = -0.07$ 
and much bluer than the source in (I-H) by $\Delta (I-H) = -0.41 \pm 0.08$. 

\section{Revisiting the modeling of the microlensing event OGLE-2012-BLG-0026 }

\subsection{A new estimate of the source star size}
H2013 converted (V-I) color to (V-K) via the Bessel and Brett relations \citep{1988PASP..100.1134B} and used the \citet{2004A&A...428..587K} V-K relation to obtain a source angular radius of $1.55 \pm 0.13 \,\mu as$. It is well known that the optical-infrared color-angular size relations
are more accurate than the ones using only optical colors \citep{2004A&A...428..587K, Boyajian:2013kh, 2014ApJ...787...92B},
and we expect that conversion from (V-I) to (V-K) colors includes this same uncertainty seen in the
optical color-angular size relations.
Using directly the \citet{2004A&A...428..587K} V-H relation, we obtain $1.58 \pm 0.10 \,\mu as$. 
Nevertheless, we decided to use the surface brightness relation from \citet{Boyajian:2013kh, 2014ApJ...787...92B} linking $\rm (V-H,~H)$ magnitudes to angular radius: 
$$\rm log(2~ \theta_* ( \mu as))= 0.536654+0.072703~(V-H)-0.2~H$$
This relation is slightly more accurate than \citet{2004A&A...428..587K} incorporating more data and excluding some 
unreliable measurements. Moreover, the fit of their relation is performed in a narrower range of spectral types providing
a better match to the source star. More details will be given by Sumi et al. (2015, submitted to ApJ).
Our revised value for the angular radius is $\rm \theta_*=1.54 \pm 0.10 \ \mu as$.  

\subsection{Modeling the photometric light curve with or without the adaptive observations constraints}
We follow the modeling approach described by \citet{2010ApJ...716.1408B} and 
\citet{2010ApJ...713..837B} for the two planet system OGLE-2006-BLG-109.
First, we take the data set presented in H2013 without alteration. We then add the CTIO H band 
photometric light curve and I band data obtained using the 0.61m B\&C telescope in New Zealand. 
In contrast with H2013, we also release the constraint on the 
distance to the source $\rm D_S$. Instead, we assume that the source is a Galactic Bulge star following
the distance distribution from the galactic model used by \citet{2014ApJ...785..155B}.
At this stage, we do not use the constraint from the measured light of the lens in H band in the modeling. 
There are 8 degenerate solutions, with 3 two-fold degeneracies in minimum impact parameter $\rm u_0$ and the 2 planet separations.
The results of the fit are given as the first column in Table 1. We obtain a slightly larger Einstein ring radius 
$\rm \theta_E=0.96 \pm 0.05~mas$ to be  compared with value derived by H2013 
of $\rm \theta_E=0.91 \pm 0.09~mas$ and a slightly larger distance to the lens and to the source, with larger error bars. 
The parallax values of the models (A, C, D) from H2013 are slightly larger, and compatible within the error bars with the results 
reported here.  
For example, the model (C and D) from H2013 give  ($\Pi_{E,N} = 0.001 \pm 0.028$, $\Pi_{E,E} = 0.123 \pm 0.005$)
and ($\Pi_{E,N} = -0.07 \pm 0.05$, $\Pi_{E,E} = 0.114 \pm 0.04$)
to be compared with ($\Pi_{E,N} = -0.004 \pm 0.03$, $\Pi_{E,E} = 0.1089 \pm 0.04$) from our unconstrained fit MOD0.

In Figure 3, we show the mass and distance of the lens star derived by H2013 and our first model MOD0,
which employs no lens brightness constraints. The MOD0 results were generated by a set Markov Chain Monte
Carlo (MCMC) over all eight of the degenerate solutions. The two results differ by more than one sigma,
but agree on the general features: the lens star is in the disk at $\approx$4~kpc, with mass in the range
 $\approx$0.8--1.1 M$_\odot$. Comparison to the theoretical stellar isochrones computed by the Padova group 
 \citep{2008A&A...484..815B} for solar metallicity and helium abundance Y=0.30 shows good consistency with a typical main-sequence 
 disk star of solar metallicity and an age of $\sim$6.4~Gyr. The 1 $\sigma$ error extending from 4 to $\sim$9~Gyr.

\begin{figure}
\includegraphics[scale=0.33,angle=0]{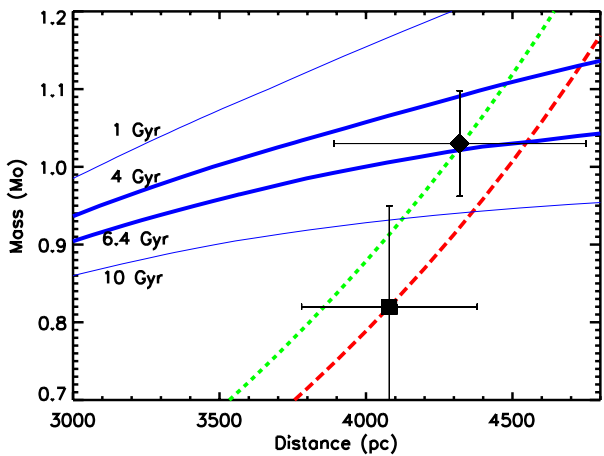}

\caption{Mass-distance relations. The H2013 solution is shown as a square, while the diamond is the first model computed in section 3.2. 
We plot in long dash red the relation $\rm M/M_\odot = \Theta_E^2 (8kpc/D_s) x / (1-x)$ for 
the parameters from H2013 and in dash green for the parameters of our MOD0 model. 
The measured H-band magnitude for the lens is $H_{L}=16.39  \pm 0.08$. We assume 
that the extinction to the lens is given by Eq.\ 1,  and we plot the mass-distance relation for the $\rm 1,~4,~6.4, ~10~Gyr$ isochrones from \citet{2008A&A...484..815B}.
\label{fig2}}
\end{figure}

We now apply the lens brightness constraints from the Keck and Subaru images to the
results of our MCMC runs used for MOD0. 
We consider two priors concerning the star formation history in the galactic disk. First a uniform star formation over 10 Gyr, 
secondly an enhanced star formation over the 4-6 Gyr period. 
This prior on the star formation history has no impact on the result. We report the solution as MOD1 in Table 1. 
Finally, we take the original MOD0 modeling, and apply the posterior Keck/Subaru constraints and star formation constraints to 
this unconstrained model, i.e., to the MOD0 results from the MCMC run. The results are identical to MOD1.
We note that the three modeling approaches yield the same central host mass, very similar values for the 
two planets, and small differences in distance to the source and to the lens. All the estimates are well within the reported $1 \sigma$ error. 

\begin{table}
\begin{center}
\caption{Physical parameters of the two-planet system OGLE-2012-BLG-0026}
\begin{tabular}{lrrr}
\multicolumn{4}{c}{ }\\
 \hline 
 \hline
Parameter & MOD0 & MOD1 & H2013 (D)\\
\hline
$\rm t_E$ ~(days)  & $94.00 \pm 0.89$ &    $94.12 \pm 0.92$        & \\
$\rm D_L$ ~(kpc)   & $4.32   \pm 0.43$ &      $4.019\pm 0.38$     &  $4.08 \pm 0.30  $ \\
$\rm D_s$~(kpc)    & $8.28   \pm 1.44$ &       $7.39 \pm  1.28$       & \\
$\rm x = D_L /D_S $        & $0.53 \pm  0.047$ & $0.55  \pm 0.05$  &\\
$\rm R_E~(AU)$            &     $ 4.12 \pm 0.41$    &  $4.02 \pm 0.38$   &\\
$\rm \theta_E~(mas)$ &     $0.96 \pm 0.048$  &  $0.98 \pm 0.04$  &  $0.91 \pm 0.09$ \\
$\rm M_*  (M_\odot$)  &  $1.06    \pm     0.05$   & $  1.06 \pm 0.05$  &       $  0.82 \pm 0.13$     \\
$\rm M_{\rm p1}$ ($\rm M_{Jup}$) &   $0.139 \pm 0.0085$       & $0.145 \pm 0.0082$ & $  0.11 \pm 0.02$ \\
$\rm M_{\rm p2}$ ($\rm M_{Jup}$) &   $0.80 \pm  0.07 $ & $0.86\pm  0.06$   & $  0.68 \pm 0.10$     \\
$\rm D_{\rm p1} (AU)$  & $3.94 \pm  0.45$ &    $4.0 \pm 0.5$   & $3.82 \pm 0.30 $   \\
$\rm D_{\rm p2} (AU)$  & $4.16 \pm  0.45$ &    $4.8 \pm 0.7$ &  $4.63  \pm 0.37$ \\
$\rm Mag_{ \rm  lens}(H)$     & $16.69  \pm 0.38$  & $16.36 \pm 0.13$ &\\
$\rm Mag_{\rm  lens}(I) $      & $19.23  \pm 0.43$  & $19.00 \pm 0.23$ &\\
$\rm Mag_{\rm lens}(V)$     & $21.75   \pm 0.50$ & $20.66 \pm 0.15$  &\\
\hline
 \end{tabular}
\end{center}
\tablecomments{We show 3 different models. MOD0 is unconstrained by the luminosity of the lens. It shows the impact of the additional lightcurve data set and estimate for the source angular size, and can directly compared to H2013 (from their table 2). MOD1 is the best model including constraints from the AO observations and Padova isochrones with uniform prior on disk star formation history. 
$\rm t_E$ is the Einstein ring crossing time, $\rm D_L$  and $\rm D_S$ the distance to the lens and to the source respectively.  $\rm R_E$  is the Einstein 
ring radius in AU, while $\rm \theta_E$ is in milliarcsec. The host star has a mass of $\rm M_*$, while the two planets are of masses $\rm M_{\rm p1}$ and $\rm M_{\rm p2}$ 
with semi major axis of $\rm D_{\rm p1}$  and $\rm D_{\rm p2}$  respectively. We provide also estimates for the V, I, H magnitudes of the lens host star.}
\end{table}

\section{Discussion and conclusion}

We revisited the microlensing event OGLE-2012-BLG-0026. We confirm the physical picture of a two-planet system of gas giants orbiting a G star. 
We improved upon previous modeling and refined the parameters of the system by adding
photometric light curves that were excluded by H2013. 
The adaptive optics observations provide further information by measuring the lens flux. 
The result is an increased estimate of the mass of the host star, to M = $1.06 \pm 0.05 M_\odot$.  
Using the H-band apparent magnitude of the lens, the reddening to the lens, and theoretical isochrones of stellar metallicity, we find good consistency for a stellar age of $\approx$4--6~Gyr.
When using all the photometric data, this microlensing event is very well constrained. When we add the Keck and Subaru constraints to the modeling, 
or apply the constraints a posteriori, it confirms our initial model, but does not allow refinement of the physical parameters of the planets. 
The physical parameters are now known to $\sim 5 \%$.  

The central star is orbited by two cold giant planets of $ \rm  0.145 \pm 0.008 \,M_{\rm Jup}$ and $ 0.86 \pm  0.07 \,M_{\rm Jup}$, at projected distances 
of $4.0 \pm 0.5\, {\rm AU}$  and $4.8 \pm 0.7\, {\rm AU}$ respectively.  Because the true orbital radii are equal to or larger than the projected separations, 
the planets are guaranteed to be well beyond the snow line for a G star. The inner planet of OGLE-2012-BLG-0026 is a roughly half the mass of Saturn, 
while the outer one is close to Jupiter. The two orbits are close. Compared to the original results, we find a more massive host star and more massive 
planets, orbiting at slightly larger projected distances.

\begin{figure}
\includegraphics[scale=0.6,angle=0]{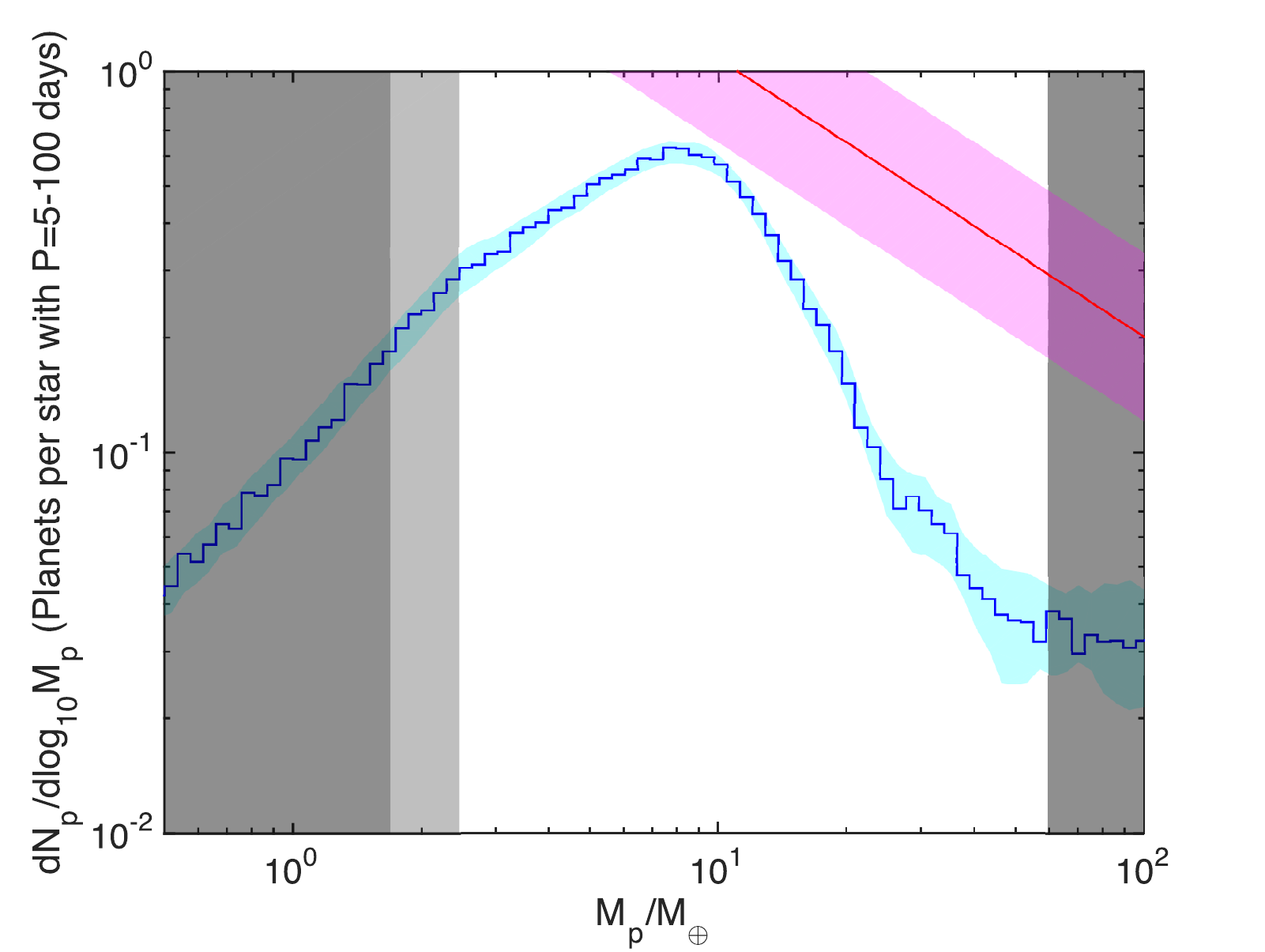}
\caption{ Mass histogram for  planets orbiting GK dwarfs over periods in the range 5-100 days discovered by Kepler. The upper left lines show the mass function
for cold planets derived by \citet{2012Natur.481..167C}. \label{fig4}}
\end{figure}

An inspection of  mass histograms  of hot and cold planets referenced in the {\it Extrasolar Planets
Encyclop{\ae}dia}\footnote{\texttt{http://exoplanet.eu}, referenced in October 2015.}  suggests that there is a dip at $ \rm \sim  45\, M_\oplus=0.141 M_{\rm Jup}$,
corresponding to the mass of the lower mass planet. However, this mass histogram is obtained by combining results from different methods 
without accounting for their detection efficiencies and biases. 

We present in Fig. 4 an estimate of the Kepler-mass distribution for quiet GK dwarf planet hosts. 
We used  the planet radius distribution from \citet{2013ApJ...770...69P}    and the \citet{2015arXiv150407557W}   probabilistic mass-radius distribution 
based on the sample of transiting planets up to $\rm \ 8 R_\oplus$ with radial velocity measurements.  
The dark grey shaded regions indicate the edges of the Mass-Radius relation being used: specifically, 
below $\rm 1.7~M_\oplus$ and above  $\rm 60~M_\oplus$ (where planets  $ \rm  \gtrsim 8 R_\oplus$) which corresponds to 
the upper limit of the  mass-radius sample on which the  Wolfgang et al. Mass-Radius relation was based.  
The pale grey region denotes the regime where incompleteness due to the small radius cut-off at $1 R_\oplus$  in the Petigura et al. radius function could start to be significant. 
The cyan shaded region accounts for the error bars quoted on the Petigura et al. radius distribution, containing 68\% of the the mass-distributions obtained
when we sampled from (gaussian-approximated) uncertainties on the planet occurrence in each radius bin. 
The y-axis of the figure is normalised such that the area under the curve gives the number of planets per star (on orbits 5-100 days) within the $\rm log(M_p)$ range chosen. 
 We remark that the planet mass distribution maybe dominated by the uncertainty in the Mass - Radius distribution, which is not reflected in the error range plotted.
We then over plot the mass function from \citet{2012Natur.481..167C} $\rm dN / (dlog(a) dlog(M) = 10^{-0.62 \pm 0.22} (M/M_{sat})^{-0.73 \pm 0.17}$ (where $M_{sat}=95 M_\oplus$).

First, we recall that the microlensing mass function is restricted to planets beyond the snow line over orbits in the range 0.5-10 AU, and spanning from $5 M_\oplus$ to $M_{\rm Jup}$ .
We have a larger abundance of cold planets, which is not surprising since it focusses on planets that most likely have 
not migrated  far from their location of formation.  Secondly, the steep slope of the microlensing mass function is coming from the statistics of \citet{2012Natur.481..167C} combining in a single power law the two populations of  gaseous giants, and super-Earth/Mini-Neptune. The results from Kepler would be suggesting that it is probably better  to separate gaseous giants, from Super-Earth/Mini Neptunes, which will be possible with the impeding increase of microlensing detections thanks to the new ground based facilities such as  the global worldwide network of 
wide field imagers  (OGLE, MOA, WISE, UTGO, KMTNet), the K2 microlensing campaign, Euclid and WFIRST. 

The physical parameters of the two planets  system  OGLE-2012-BLG-0026 are constrained to 5 \%. Both planets are typical gaseous planets orbiting a solar like star. 
We also have derived a constraint on its age, corresponding to the standard age of disk stars for a solar metallicity.
Given the absence of blends at the arcsecond scale and the fact that the lens is brighter than the source, it will be possible in the future to directly measure the metallicity with
spectroscopy using 8m class telescopes.




\acknowledgments

V.B. was supported by the CNES and the DIM ACAV, R\'egion \^Ile-de-France. 
V.B., J.P.B., and J.B.M.  acknowledge the support of PERSU Sorbonne Universit\'e, the Programme National de Plan\'etologie and the labex ESEP.
We are grateful to F. Naudin for discussions about the properties of this system.
D.P.B. was supported by grants NASA-NNX12AF54G, JPL-RSA 1453175 and NSF AST-1211875. F.A. is supported by JSPS23340064
Work by C.H. was supported by Creative Research Initiative Program (2009-0081561) of National Research Foundation of Korea.
This work was partially supported by a NASA Keck PI Data Award, administered by the NASA Exoplanet Science
Institute. Data presented herein were obtained at the W.M. Keck Observatory from telescope time
allocated to the National Aeronautics and Space Administration through the agency scientific
partnership with the California Institute of Technology and the University of California. The
Observatory was made possible by the generous financial support of the W.M. Keck Foundation.
This work was performed [in part] under contract with the Jet Propulsion Laboratory (JPL) funded by NASA through the Sagan 
Fellowship Program executed by the NASA Exoplanet Science Institute. 
Work by J.C.Y. was performed under contract with the California Institute of Technology (Caltech) / Jet
Propulsion Laboratory (JPL) funded by NASA through the Sagan Fellowship Program executed
by the NASA Exoplanet Science Institute. This research made use of \texttt{astropy}, a community-developed core Python package for Astronomy, \texttt{astroML}, and \texttt{TOPCAT}. This publication makes use of data products from the Two Micron All Sky Survey, which is a joint project of the University of Massachusetts and the Infrared Processing and Analysis Center/California Institute of Technology, 
funded by the National Aeronautics and Space Administration and the National Science Foundation.
Based on data products from observations made with ESO Telescopes at the La Silla or Paranal Observatories 
under ESO programme ID 179.B-2002.




\clearpage

\clearpage

\bibliographystyle{apj}
\bibliography{ms}

\end{document}